\documentclass[twocolumn,pre,showpacs,showkeys]{revtex4}
\usepackage{hyperref}
\usepackage{amsmath}
\usepackage{bm}
\usepackage{graphicx}
\begin{document}
\title{Steady-state stabilization due to random delays in maps with
self-feedback loops and in globally delayed-coupled maps}

\author {Arturo C. Mart\'{\i}$^a$}
%\email{marti@fisica.edu.uy}
\author{ Marcelo Ponce$^a$}
%\email{mponce@fisica.edu.uy},
\author{C. Masoller$^{a,b}$}
%\email{cris@fisica.edu.uy}

\affiliation{$^a$Instituto de F\'{\i}sica, Facultad de
  Ciencias, Universidad de la Rep\'ublica, Igu\'a 4225, Montevideo
  11400, Uruguay} \affiliation{$^b$Departament de Fisica i Enginyeria
  Nuclear, Universitat Politecnica de Catalunya, Colom 11, E-08222
  Terrassa, Spain}

\date{\today}

\begin{abstract}
We study the stability of the fixed-point solution of an array of
mutually coupled logistic maps, focusing on the influence of the
delay times, $\tau_{ij}$, of the interaction between the $i$th and
$j$th maps. Two of us recently reported [Phys. Rev. Lett. {\bf 94},
134102 (2005)] that if $\tau_{ij}$ are random enough the array
synchronizes in a spatially homogeneous steady state. Here we study
this behavior by comparing the dynamics of a map of an array of $N$
delayed-coupled maps with the dynamics of a map with $N$
self-feedback delayed loops. If $N$ is sufficiently large, the
dynamics of a map of the array is similar to the dynamics of a map
with self-feedback loops with the same delay times. Several delayed
loops stabilize the fixed point, when the delays are not the same;
however, the distribution of delays plays a key role: if the delays
are all odd a periodic orbit (and not the fixed point) is
stabilized. We present a linear stability analysis and apply some
mathematical theorems that explain the numerical results.
\end{abstract}

\pacs{05.45.Xt, 05.65.+b, 05.45.Ra} \keywords{Random delays;
synchronization; coupled maps; logistic map}

\maketitle

\section{Introduction}
Cooperative behavior arises in many fields of science and classical
examples include the onset of rhythmic activity in the brain, the
flashing on and off in unison of populations of fireflies, and the
emission of chirps by populations of crickets \cite{review}. Of
important practical applications are the synchronization of laser
arrays and Josephson junctions. Coupled map lattices (CMLs)
\cite{kaneko} are excellent tools for understanding the mechanisms
of emergency of synchrony in complex systems composed of interacting
nonlinear units, because, by simplifying the dynamics of the
individual units, CMLs allow the simulation of large ensembles of
coupled units.

The effects of time delays arising from the finite propagation time
of signals have received considerable attention. Classical examples
are the Mackey-Glass model in physiology \cite{mackey} (that
describes anomalies in the regeneration of white blood cells due to
the finite time of propagation of chemical substances in the blood),
and the Ikeda model \cite{ikeda} in optics (that accounts for the
finite velocity of light in optical bistable devices).

Three common consequences of time-delays are multistability, which
typically arises for delays longer than the intrinsic oscillation
period \cite{longtin,yeung}, chaotic dynamics, which arises for
strong coupling and/or long delays, and oscillation death, which
refers to the existence of stability islands in the parameter space
(coupling strength, delay time) where the amplitude of coupled
limit-cycle oscillators is zero \cite{od}. It is also well-known
that time-delayed feedback can stabilize unstable orbits embedded in
chaotic attractors \cite{pyragas} and enhance the coherence of
chaotic \cite{coherence,boccaletti} and stochastic motions
\cite{scholl}.

Most studies of delayed coupling have considered uniform delays,
i.e., the interactions between the different elements of a network
occur all with the same delay time (instantaneous coupling is a
particular case of ``fixed delay coupling''). To the best of our
knowledge, the first study of a system of mutually coupled units
interacting with different, randomly chosen delay times was done by
Otsuka and Chern \cite{otsuka} in the early 90's. In Ref.
\cite{otsuka} an array of semiconductor lasers with incoherent
optical coupling was studied numerically, and it was shown that
different dynamic regimes occur, including synchronization,
clustering and steady-state behavior, depending on the average delay
and the delay distribution. Distributed (or random) delays have been
the object of recent attention since several authors have reported
that non-uniform delays can have a stabilizing effect. Atay et al.
\cite{atay_PRL_2003} studied ensembles of limit-cycle oscillators
and showed that distributed delays can enlarge the stability islands
where oscillator death occurs. Huber and Tsimring \cite{lev} studied
networks of globally coupled, noise-activated, bistable elements and
found that increasing the non-uniformity of the delays enhanced the
stability of the trivial equilibrium. Eurich et al. \cite{eurich}
showed that distributed delays increase the stability of
predator-prey systems including two-species systems, food chains,
and food webs.

Two of us recently studied an array of logistic maps coupled with
randomly distributed delay times \cite{prl},
\begin{equation}
\label{array}
 x_i(t+1)= f[x_i(t)] + \frac {\epsilon} N  \sum_{j=1}^N \left(
f[x_j(t-\tau_{ij})] - f[x_i(t)] \right),
\end{equation}
where $t\ge 0$ is an integer-valued time index, $i=1,\dots,N$ is a
space index, $f(x)=ax(1-x)$ is the logistic map ($a \in (0,4]$),
$\epsilon$ is the coupling strength ($\epsilon \in [0,1]$) and
$\tau_{ij} \ge 0$ is an integer that represents the delay time in
the interaction between the $i$th and $j$th maps. For $\tau_{ij}$
random enough the array synchronizes in the spatially homogeneous
steady-state, $x_i(t)=x_0$ for all $i$, where $x_0$ is the
nontrivial fixed point ($x_0=1-1/a$). This synchronization behavior
is in contrast with the synchronization with fixed and
distant-dependent delays. For fixed delays ($\tau_{ij}=\tau_0$
$\forall$ $i$, $j$) the array synchronizes in a spatially
homogeneous time-dependent state, $x_i(t)=x(t)$ $\forall$ $i$, $t$,
where the dynamics of an element of the array is either periodic or
chaotic depending on $\tau_0$ \cite{atay_PRL_2004}. For
distant-dependent delays ($\tau_{ij}=k|i-j|$ where $k$ is the
inverse of the velocity of transmission of information) a
one-dimensional linear array synchronizes in a state in which the
elements of the array evolve along a periodic orbit of the uncoupled
map (i.e., $x_i(t)$ is a solution of $x_i(t+1)=f[x_i(t)]$), while
the spatial correlation along the array is such that $x_i(t) =
x_j(t-\tau_{ij})$ $\forall$ $i$, $j$ (i.e., a map sees all other
maps in his present, current, state) \cite{marti}.

In Ref.\cite{prl} the stabilization of the fixed-point solution due
to random interaction delay times was interpreted as a ``discrete
time'' version of the control method for stabilizing a fixed point
recently proposed by Ahlborn and Parlitz \cite{parlitz}. In
Ref.\cite{parlitz} the fixed point of a dynamical system
$\dot{x}=f(x)$ was stabilized with the addition of several feedback
terms that satisfy: (i) the feedback terms vanish in the steady
state and (ii) the delay times are not an integer multiple of each
other (with these conditions the control terms vanish only at the
fixed points and not at the periodic orbits). Simulations of a
single logistic map with $N$ self-feedback time-delayed loops,
\begin{equation}
\label{un_mapa}
 x(t+1)= f[x(t)] + \frac {\epsilon} N\sum_{j=1}^N
\left(f[x(t-\tau_{j})] - f[x(t)]\right),
\end{equation}
show that several terms with different delays lead to the
stabilization of the fixed point after transients.

The behavior of a single unit often helps understanding the behavior
of an ensemble of coupled units, and in particular the "chaos
suppression by random delays" in an ensemble of coupled logistic
maps can be interpreted in terms of the suppression of chaos and the
stabilization of the fixed point in a single logistic map with
several delayed self-feedback loops. The aim of this paper is to
further investigate this point, by comparing the dynamics of an
element $x_i$ of an array of globally coupled $N$ logistic maps
[Eq.~(\ref{array})] with the dynamics of a logistic map with $N$
self-feedback loops [Eq.~(\ref{un_mapa})].

This paper is organized as follows. Section II presents a linear
stability analysis of the fixed point solution of
Eq.~(\ref{un_mapa}) and discusses the stability in the parameter
space (local nonlinearity, $a$, feedback strength, $\epsilon$). We
find an analytical (sufficient) instability condition,
Eq.(\ref{inestable_seguro}), that holds for large $a$ and low
$\epsilon$, regardless of the number of feedback terms and/or the
values of the delay times. We find a second (sufficient) instability
condition, Eq.(\ref{inestable_seguro2}), that holds for one feedback
loop, independently of the delay time. Section III presents a
comparison of the dynamics of a logistic map with $N$ self-feedback
delayed loops, with the dynamics of a map of an array of $N$
delayed-coupled logistic maps. The numerical simulations show that
if $N$ is sufficiently large, the dynamics of a map of the array is
remarkably similar to the dynamics of a map with $N$ feedback loops
and the same delay times. The similarities are explored by analyzing
the regions in the parameter space ($a$, $\epsilon$) where the
fixed-point solution is stable, and comparing with the analytic
results of Sec. II. We also present bifurcation diagrams that
demonstrate similar types of instability scenarios. Section IV
presents an interpretation of these results based on the analogy
between globally coupled maps and a single map with external
driving, studied by Parravano and Cosenza in
Refs.\cite{cosenza1,cosenza2}. Section V presents a summary and the
conclusions.

\section{Stability Analysis}

To analyze the stability of the nontrivial fixed point solution of
Eq.~(\ref{un_mapa}), $x_0=f(x_0)$ (for the logistic map $x_0=1-1/a$),
we define a new set of variables,
\begin{equation}
y_{m}(t)=x(t-m),
\end{equation}
with $m=0, ..., M$ and $M=\max(\tau_{j})$, that describe the present
and past state of the map. We can re-write Eq.(\ref{un_mapa}) in
terms of these new variables as:
\begin{eqnarray}
y_m(t+1)=
\begin{cases}
              y_{m-1}(t)& \text{if $m \ne 0$},\\
              (1-\epsilon)f[y_{0}(t)]+\frac{\epsilon}{N}
              \sum_{i=1}^N f[y_{ \tau_{i}}] & \text{if $m=0$}.
              \end{cases}
\end{eqnarray}

The fixed-point solution is
\begin{equation}
y_0(t)=x_0; \dots ; y_M(t)=x_0.
\end{equation}

To study the stability of this solution we linearize,
\begin{equation}
\delta y_{m} (t+1) = \sum_{n=0}^M A_{mn} \delta y_{n} (t),
\end{equation}
where
\begin{equation}
A=
\begin{pmatrix}
(1-\epsilon)f'(x_{0})+\alpha_0 & \alpha_{1}  & \alpha_{2} & \hdots &
\alpha_{M-1}&
 \alpha_{M} \\
 1 & 0 & 0 & \hdots &0  & 0 \\
0 & 1  &0 & \hdots &0 & 0 \\
\vdots & & & &\vdots \\
0 & 0& 0& \hdots &0 & 0 \\
0 & 0& 0& \hdots & 1 & 0
\end{pmatrix}.
\end{equation}
Here
\begin{equation}
\alpha_{n}= \frac \epsilon N
              \sum_{i=1}^N f'[y_{\tau_i}(t)]\delta_{\tau_i n}
= l_n \frac \epsilon N
               f'(x_0),
\label{suma}
\end{equation}
where $l_n$ is the number of times the value $\tau=n$ appears in the
sequence $\tau_1$, $\tau_2$,... $\tau_N$: $\sum_{n=1}^M l_n=N$.
%For simplicity we assume that $\tau_j \ne 0$ $\forall$ $j$ (the analysis
%can be extended to account for instantaneous feedback loops).
The term $\alpha_0$ accounts for the instantaneous feedback loops.
Notice that some of the $\alpha_{n}$ coefficients will be zero
($\alpha_{n}=0$ if $\tau_i\ne n$ $\forall i$); however, the
coefficient corresponding to the maximum delay, $M=\tau_{max}$, is
different from zero and is given by
 \begin{equation}
\alpha_{M}=  l_M \frac \epsilon N f'(x_0).
\end{equation}
The next step for the derivation of analytic stability conditions is
the study the eigenvalues $\lambda_i$ (with $i=0\dots M$) of the
matrix $A$. The Gershgorin theorem \cite{teorema} states that all
eigenvalues of a complex square matrix are located in a set of disks
centered at the diagonal elements $a_{ii}$ with radius equal to the
sum of the norms of the other elements on the same row:
\begin{eqnarray}
\label{gershgorin} |\lambda_i - a_{ii}| &\le& \sum_{j\ne i}
|a_{ij}|.
\end{eqnarray}
For $i\ne0$ Eq.~(\ref{gershgorin}) gives $|\lambda_i| \le 1$ and for
$i=0$ gives
\begin{eqnarray}
|\lambda_i - (1-\epsilon)f'(x_{0})-\alpha_0| &\le&
%\sum_{j=1}^M |\alpha_{j}|
 \epsilon |f'(x_0)| - | \alpha_0 |,
\end{eqnarray}
where we used Eq.~(\ref{suma}), $\sum_{j=0}^M
 l_j=N$ and $\sum_{j=0}^M \alpha_{j}=\epsilon
f'(x_{0})$. Therefore, the eigenvalues are in the region of the
complex plane defined by the two disks:
\begin{eqnarray}
|\lambda| &\le& 1 \\
|\lambda - (1-\epsilon)f'(x_{0})- \alpha_0| &\le& \epsilon |f'(x_0)|
- |\alpha_0| .
\end{eqnarray}
From here we derive a sufficient stability condition and a
sufficient instability condition. If the disc of radius $\epsilon
|f'(x_0)|- |\alpha_0|$ centered at $(1-\epsilon)f'(x_{0})+ \alpha_0$
is completely inside the disc of radius 1 centered at 0, then all
the eigenvalues will have $|\lambda|<1$. Therefore, a sufficient
stability condition is $|(1-\epsilon)f'(x_{0})+\alpha_0|+\epsilon
|f'(x_0)|-|\alpha_0|<1$, and taking into account Eq.~(\ref{suma})
all the $\alpha_n$ and $f'(x_0)$ have the same sign, the sufficient
stability  condition read as
\begin{equation}
\label{estable_seguro}
|f'(x_0)|<1.
\end{equation}
This stability condition is trivial because if
Eq.(\ref{estable_seguro}) holds, then the fixed point of the
``solitary map'' (the map without feedback loops, $\epsilon=0$) is
stable.

Let us consider the region where $|f^{'}(x_0)|>1$ (for the logistic
map $|f^{'}(x_0)|>1$ for $a>3$). In this region we have the
following sufficient instability condition: if
\begin{eqnarray}
\label{inestable_seguro}
 |(1-\epsilon)f'(x_{0})|-\epsilon |f'(x_0)|+ 2 |\alpha_0| &>&1,
\end{eqnarray}
 then the disc centered at $(1-\epsilon)f'(x_{0})+ \alpha_0$ of
radius $\epsilon |f'(x_0)|- |\alpha_0|$ is completely outside of the
disc centered at zero of radius 1, and therefore there is at least
one eigenvalue with $|\lambda| > 1$. We remark that this instability
condition holds, regardless of the number of feedback terms, and
regardless the values of the delay times.

The stability and instability regions are displayed in Fig. 1(a).
The trivial stability condition holds for $a \le 3$, the instability
condition holds for large $a$ and small $\epsilon$ (there is a
second instability region, in the corner of large $a$ and $\epsilon$
which is discussed below). For comparison, we show in Figs.
1(b)-1(d) the stability regions calculated from numerical
simulations of Eq.(\ref{un_mapa}) with one delayed feedback loop
(and different delay times), which agree well with the analytic
predictions.
\begin{figure}
\center
 \resizebox{0.8\columnwidth}{!}{\includegraphics{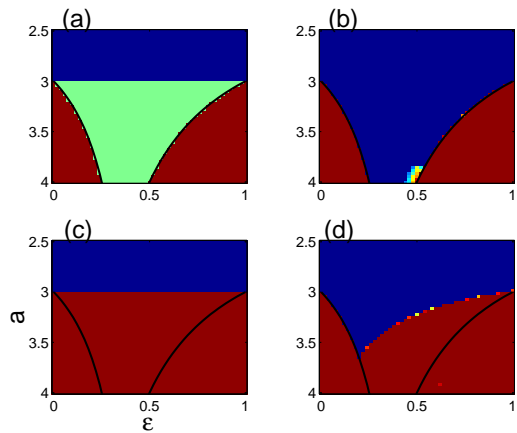}}
\caption{(a) Analytically calculated stability regions for the
logistic map with one delay term. The trivial stability region
defined by Eq. (\ref{estable_seguro}) is displayed in black (blue
online), the instability regions defined by Eqs.
(\ref{inestable_seguro}) and (\ref{inestable_seguro2}) are displayed
in dark grey (red online). (b)-(d) Stability regions calculated
numerically, by simulation of Eq. (\ref{un_mapa}) with one delay
term and different initial conditions. The parameter regions where
the fixed point is stable for all initial conditions are displayed
in black (blue online). The black solid lines indicate the borders
of the instability regions defined by Eqs.(\ref{inestable_seguro})
and (\ref{inestable_seguro2}). Near these boundaries there is
sensitivity to the initial conditions: some trajectories evolve
towards the fixed point, while others evolve to periodic or chaotic
orbits. The delay times are (b) $\tau_1=1$, (c) $\tau_1=2$, (d)
$\tau_1=3$.}
\end{figure}

Additional information can be obtained by calculating explicitly the
eigenvalues of $A$. The roots of the characteristic equation
\begin{equation}
\det \begin{pmatrix} (1-\epsilon)f'(x_{0})+ \alpha_0 -\lambda & \alpha_{1} &
\alpha_{2} & \hdots & \alpha_{M-1}&
 \alpha_{M} \\
 1 & -\lambda & 0 & \hdots &0  & 0 \\
0 & 1  &-\lambda & \hdots &0 & 0 \\
\vdots & & & &\vdots \\
0 & 0& 0& \hdots &-\lambda & 0 \\
0 & 0& 0& \hdots & 1 & -\lambda
\end{pmatrix}=0,
\end{equation}
can be written in terms of the determinant of two $M\times M$
matrices:
\begin{eqnarray}
\label{a0}
 [(1-\epsilon)f'(x_{0})+ \alpha_0 -\lambda] \det C - \det B=0,
\end{eqnarray}
where
\begin{eqnarray}
C=
\begin{pmatrix}
  -\lambda & 0 & \hdots &0  & 0 \\
 1  &-\lambda & \hdots &0 & 0 \\
\vdots & & & &\vdots \\
 0& 0& \hdots &-\lambda & 0 \\
 0& 0& \hdots & 1 & -\lambda
\end{pmatrix}
\end{eqnarray}
and
\begin{eqnarray}
B=
\begin{pmatrix}
\alpha_{1}  & \alpha_{2} & \hdots &
\alpha_{M-1}&
 \alpha_{M} \\
 1  &-\lambda & \hdots &0 & 0 \\
\vdots & & & &\vdots \\
 0& 0& \hdots &-\lambda & 0 \\
 0& 0& \hdots & 1 & -\lambda
\end{pmatrix}.
\end{eqnarray}
We calculated the determinant of each matrix recursively (details
are presented in the appendix) obtaining:
\begin{eqnarray}
\label{polinomio} \lambda^{M+1} - [(1-\epsilon) f'(x_{0})+ \alpha_0
]\lambda^M - \sum_{j=1}^M \alpha_{j}\lambda^{M-j} =0.
\end{eqnarray}
The roots of Eq.(\ref{polinomio}) satisfy $\prod_{i=0}^{M}
|\lambda_i| = |\alpha_M|$. Thus, if
\begin{equation}
\label{inestable_seguro2}
|\alpha_M|=|l_M \epsilon f'(x_0)/N|>1
\end{equation}
at least one eigenvalue has $|\lambda|>1$, i.e., this gives another
analytic (sufficient) instability condition. For $l_M/N=1$
Eq.(\ref{inestable_seguro2}) holds in the right-bottom corner of
Fig. 1(a) (large $a$ , large $\epsilon$); for $l_M/N<1$
Eq.(\ref{inestable_seguro2}) is not satisfied in the parameter
region of interest ($a$ $\in$ [0,4], $\epsilon$ $\in$ [0,1]). Notice
that if $l_M/N=1$, then the map has only one feedback loop (because
the multiplicity of the feedback terms with maximum delay is equal
to the total number of feedback terms); therefore,
Eq.(\ref{inestable_seguro2}) indicates that when a logistic map has
a single feedback term, the fixed-point solution can not be stable
in the parameter region (large $a$, large $\epsilon$), regardless of
the delay time. In other words, a single feedback term can not
stabilize the fixed point in the (large $a$, large $\epsilon$)
parameter region. Numerical simulations of Eq.(\ref{un_mapa}) with
$N=1$ and different delays confirm these analytical predictions, see
Figs. 1(b)-(d).

Further analytical insight can be gained by considering two special
cases: all-even and all-odds delays. First, let us show that if the
delays are all even (and therefore, $M$ is even), $\lambda=-1$ is a
solution of Eq. (\ref{polinomio}) when $f'(x_{0})=-1$, i.e., at the
border of the stability region, Eq.(\ref{estable_seguro}).
Substituting $\lambda=-1$ in Eq.(\ref{polinomio}) and taking into
account that $M-j$ is even (since in the sum only terms with $j$
even are different from zero) gives
\begin{eqnarray}
 -1 - [(1-\epsilon) f'(x_{0})+ \alpha_0] - \sum_{j=1}^M \alpha_{j} =0.
\end{eqnarray}
Using $\sum_{j=0}^M \alpha_{j}=\epsilon f'(x_{0})$ we obtain
$f'(x_{0})=-1$. Therefore, when the delays are all-even there is an
eigenvalue $\lambda=-1$ if and only if $a=3$, regardless of
$\epsilon$.

Next, let's see what happens if the delay times are all odd,
therefore, $M$ is odd and, in addition, $\alpha_0=0$. Taking into
account that $M-j$ is even (since in the sum only terms with $j$ odd
are different from zero) for $\lambda=-1$ Eq. (\ref{polinomio})
gives
\begin{eqnarray}
1 + [(1-\epsilon) f'(x_{0})] - \sum_{j=1}^M \alpha_{j} =0.
\end{eqnarray}
Using $\sum_{j=1}^M \alpha_{j}=\epsilon f'(x_{0})$ we obtain
$f'(x_{0})=-1/(1-2\epsilon)$. For the logistic map this gives
$a=(3-4\epsilon)/(1-2\epsilon)$ a condition which is satisfied for
$\epsilon \in [0,1]$ only for values of $a$ inside the stability
region $a<3$.

The above analysis allows as to draw some additional conclusions
about the stability of the fixed point in the special cases of
all-even and all-odd delays. For all-even delays an eigenvalue is
real and negative and equal to $-1$ for $a=3$. For larger $a$ this
eigenvalue can in principle become $\lambda<-1$ rendering the fixed
point unstable due to a period-doubling bifurcation. For all-odd
delays this instability scenario is not possible, as $\lambda=-1$ in
a parameter region where we know that all eigenvalues must have
$|\lambda| \le 1$.

We verified these predictions by calculating numerically the
eigenvalues of $A$. Figure 2 displays results for all-odd and
all-even delays, varying $a$ while keeping $\epsilon$ constant. It
can be observed that for all-even delays one real eigenvalue becomes
less than $-1$ for $a>3$; for all-odd delays a pair of
complex-conjugate eigenvalues have modulus greater than $1$ for
$a>3.8$.
\begin{figure}
\center
\resizebox{0.80\columnwidth}{!}{\includegraphics{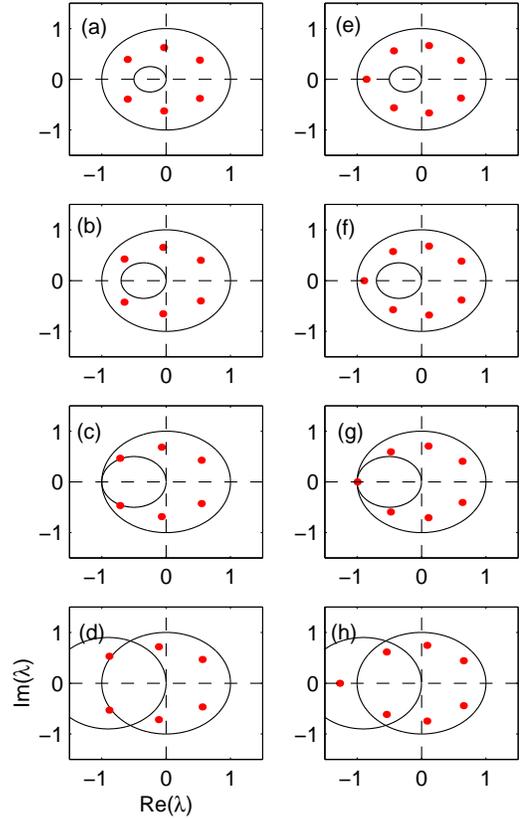}}
\caption{Eigenvalues in the complex plane for a map with $N=3$
feedback loops with all-odd delays (left column: $\tau_j=$1, 3, 5)
and all-even delays (right column: $\tau_j=$2, 4, 6). $\epsilon=0.5$
and (a), (e) $a=2.5$; (b), (f) $a=2.7$; (c), (g) $a=3.0$ and (d),
(h) $a=3.8$. The circles indicate the Gershgorin disks.}
\end{figure}

\section{Numerical Results}

In this section we compare the dynamics of a logistic map with $N$
self-feedback loops, Eq. (\ref{un_mapa}), with the dynamics of a map
of an array of $N$ globally delayed-coupled logistic maps, Eq.
(\ref{array}).

We consider Gaussian distributed delays: $\tau_{ij} = \tau_0 +
\mathrm{Near} (c \xi)$, where $c$ is a parameter that allows varying
the width of the delay distribution (for $c=0$ the delays are all
equal, $\tau_{ij}=\tau_0$, for $c\ne 0$ the delays are distributed
around $\tau_0$); $\xi$ is Gaussian distributed with zero mean and
standard deviation one; $\mathrm{Near}$ denotes the nearest integer
(we use $\mathrm{Near}$ instead of $\mathrm{Int}$ to have a
distribution that is symmetric with respect to $\tau_0$; however,
the results are largely independent of the precise form of the delay
distribution). Depending on $\tau_0$ and $c$ the distribution of
delays has to be truncated to avoid negative delays.

We begin by showing that if $N$ is large enough, the parameter
region where the fixed point is stable for the array is remarkably
similar to the parameter region where the fixed point is stable for
the map with feedback loops. Figure 3 displays results for three
values of $N$, the upper row shows the stability region of the
homogeneous steady-state solution of the array [$x_i(t)=x_j(t)=x_0$
$\forall$ $i$, $j$, $t$], while the lower row shows the stability
region of the fixed point solution the map with feedback loops. The
delays of the self-feedback terms, $\tau_j$ with $j=1\dots N$, were
taken equal to the delay times of the interaction of the $i$th and
$j$th maps of the array; this gives $N$ sets of delay times
($\tau_j=\tau_{ij}$ with $i=1\dots N$). In the lower row of Fig. 3,
the parameter region where the fixed point is stable for all sets of
delays is displayed in black (blue online), and the regions where is
unstable for all sets of delays are displayed in dark gray (red
online). Outside the trivial stability region ($a\le3$) and outside
the instability region defined by the sufficient condition Eq.
(\ref{inestable_seguro}), if $N$ is small the fixed point of the map
with self-feedback loops can be stable or unstable depending on
$\tau_j$ (i.e., the fixed point can be stable for the $i$th set of
delays and not for the $k$th set of delays); however, if $N$ is
sufficiently large the stability of the fixed point is the same for
all sets of delays (there is sensitivity to the precise values of
$\tau_j$ near the boundaries of the instability regions). It can be
observed that for both, the single map and the array, the fixed
point is unstable in the left-bottom corner (large $a$, low
$\epsilon$), in agreement with the results of the previous section
(where we showed that in this region the sufficient instability
condition, Eq.(\ref{inestable_seguro}), holds).
\begin{figure}
\center
 \resizebox{1.0\columnwidth}{!}{\includegraphics{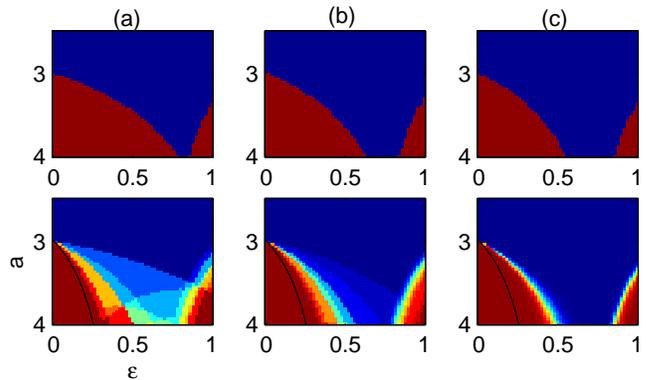}}
\caption{Stability region of the fixed point solution of an array of
$N$ maps (upper row) and of a map with $N$ feedback loops (lower
row). The delays are Gaussian distributed with $\tau_0=3$, $c=1$.
The dynamics of the map with feedback loops was simulated for
various sets of delays $\tau_j$: the fixed point was found to be
stable for all sets in the black region (blue online) and unstable
for all sets in the dark gray regions (red online). The solid line
indicates the borders of the instability region defined by Eq.
(\ref{inestable_seguro}). (a) $N=10$, (b) $N=20$, (c) $N=100$.}
\end{figure}

The stability of the fixed point depends on the distribution of
delays, and again, the similarities between a map of the array and a
map with self-feedback loops, for $N$ large enough, are remarkable.
As the width of the delay distribution, $c$, increases, the
parameter region where the fixed point is stable grows (see Fig. 4),
and this occurs for both, the array and the map with self-feedback
loops.
\begin{figure}
\center
 \resizebox{1.0\columnwidth}{!}{\includegraphics{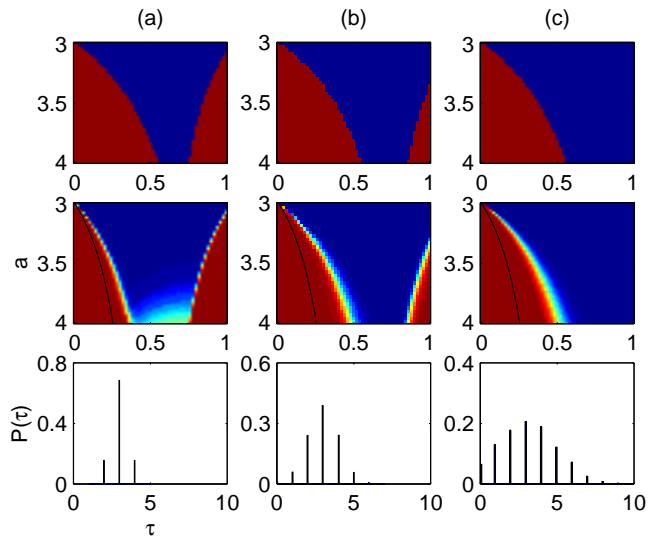}}
\caption{Influence of the width of the delay distribution on the
stability of the fixed point solution of an array of $N$ maps (upper
row) and of a map with $N$ feedback loops (middle row). The delays
(lower row) are Gaussian distributed with $\tau_0=3$, (a) $c=0.5$,
(b) $c=1$ , (c) $c=2$. $N=100$. }
\end{figure}

For the array of coupled maps, the parameter that quantifies the
influence of the delays is not the mean delay, $<\tau>$, or the
standard deviation of the distribution, $D_\tau$, but is the
normalized disorder parameter, $c^\ast=D_\tau/<\tau>$ \cite{prl}
($<\tau>=\tau_0$, $D_\tau=c$, and $c^\ast=c/\tau_0$ if the Gaussian
distribution is not truncated). Figure 5 shows that this is also the
case for the map with self-feedback loops, as it can be noticed that
the stability region of the fixed point is the same for
distributions that have different $<\tau>$ and $D_\tau$, but the
same normalized width, $c^\ast$.
\begin{figure}
\center
 \resizebox{1.0\columnwidth}{!}{\includegraphics{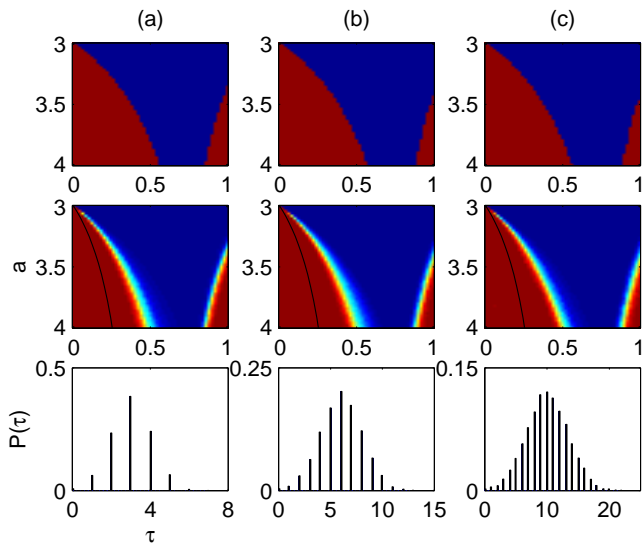}}
\caption{Influence of the normalized disorder parameter on the
stability of the fixed point solution of an array of $N$ maps (upper
row) and of a map with $N$ feedback loops (middle row). The delays
(lower row) are Gaussian distributed with different mean and
different standard deviation but with the same normalized width,
$c^\ast \sim c/\tau_0$ (a) $\tau_0=3$, $c=1$, (b) $\tau_0=6$, $c=2$
, (c) $\tau_0=10$, $c=3.33$. $N=100$.}
\end{figure}

Figure 6(a) displays the stability region of the fixed point
solution for the array (upper row) and for the map with
self-feedback loops (lower row), in the parameter space [$\tau_0$
($\sim <\tau>$), $c$ ($\sim D_\tau$)]. It can be observed that the
fixed point is stable if the delays are random enough (i.e., if the
width of the distribution, $c$, is larger than a certain value that
increases with $\tau_0$). When the stability region is plotted vs.
the normalized width, $c^\ast$ [Fig. 6(b)], it can be observed than
the value of $c^\ast$ above which the fixed point is stable is
independent of $\tau_0$ [but depends on $\epsilon$, as shown in Fig.
6(c)]. This occurs for both, the map of the array and the map with
self-feedback loops.
\begin{figure}
\center
 \resizebox{1.0\columnwidth}{!}{\includegraphics{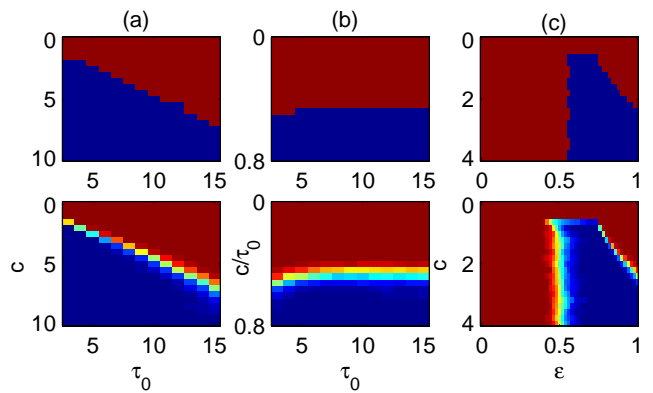}}
\caption{(a) Stability region of the fixed point solution of an
array of $N$ maps (upper row) and of a map with $N$ feedback loops
(lower row) in the parameter space (mean delay, standard deviation).
(b) Same as (a) but plotted vs. the normalized disorder parameter,
see text. (c) Same as (a) but in parameter space ($\epsilon$,
disorder parameter). $N=100$, $a=4$, in (a),(b) $\epsilon=1$; in (c)
$\tau_0=5$.}
\end{figure}

While the above observation of enhanced fixed-point stability with
increasing randomness of delays is generic, independent of the
precise form of the delay distribution, there is an exception which
is the case of all-even delays. For all-even delays the fixed point
of the map with self-feedback terms is stable only in the region
defined by the sufficient trivial stability condition,
Eq.(\ref{estable_seguro}). Because the fixed point becomes unstable
due to a period-doubling bifurcation when one real eigenvalue
becomes $\lambda<-1$ (as discussed in the previous section),
all-even delays tend to stabilize an orbit of period 2. The same
effect is observed in the array of $N$ logistic maps: for all even
delays the fixed point is stable only in the trivial region $a \le
3$.

If $N$ is sufficiently large, a map with $N$ self-feedback loops and
a map of an array of $N$ maps follow very similar instabilities
scenarios when $\epsilon$ or $a$ are varied (see below for a
discussion of the limit in which the similarities are not only
qualitative but also quantitative). As an example, Figs. 7-9 display
bifurcation diagrams for varying $\epsilon$ while keeping $a$ fixed.
The delays are ``mixed'' (even and odd) in Fig. 7, all-odd in Fig.
8, and all-even in Fig. 9. The parameters correspond to a scan of
$\epsilon$ along the horizontal axis of Fig. 4(b): for ``mixed''
delays the fixed point is stable in a range of $\epsilon$ for both,
the map with feedback loops and the array. Figures 7(a)-9(a)
[7(b)-9(b)] display the time evolution of one element of the array,
$i=1$ ($i=2$), by plotting 100 consecutive interactions of $x_1$
($x_2$) after transients die away vs. $\epsilon$. Figures 7(c)-9(c)
display the array configuration at time $t=T$ (large enough to let
transients die away). Figures 7(d)-9(d) [7(e)-9(e)] display the time
evolution of the map with feedback loops with delays
$\tau_j=\tau_{1j}$ ($\tau_j=\tau_{2j}$), by plotting 100 consecutive
interactions of $x$ (after transients die away) vs. $\epsilon$.
\begin{figure}
\center
 \resizebox{1.0\columnwidth}{!}{\includegraphics{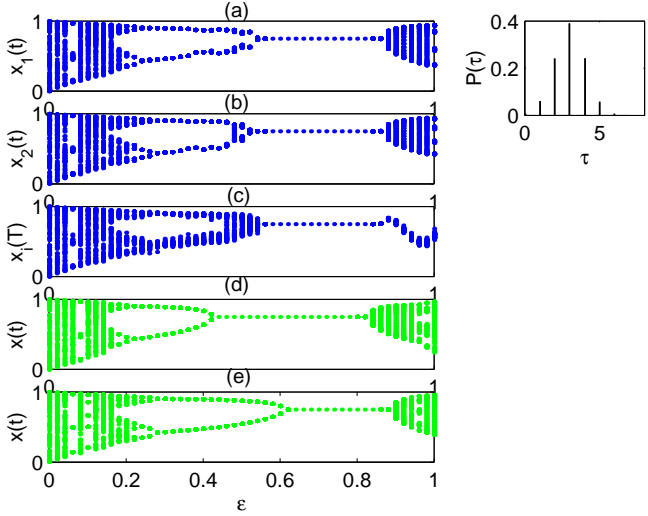}}
\caption{Bifurcation diagrams for increasing $\epsilon$. (a) $i=1$
map of the array. (b) $i=2$ map of the array. (c) Array
configuration, $x_i$ with $i=1,\dots,N$ at time $t=T$. (d) A map
with feedback loops with delays $\tau_j=\tau_{1j}$. (e) A map with
feedback loops with delays $\tau_j=\tau_{2j}$. $a=4$, $N=100$, the
delays are Gaussian distributed with $\tau_0=3$, $c=1$.}
\end{figure}
\begin{figure}
\center
 \resizebox{1.0\columnwidth}{!}{\includegraphics{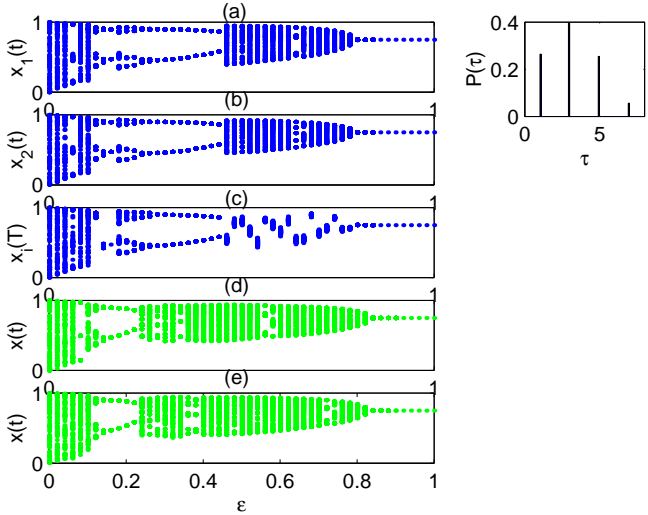}}
\caption{As Fig. 7 but with all-odd delays.}
\end{figure}
\begin{figure}
\center
 \resizebox{1.0\columnwidth}{!}{\includegraphics{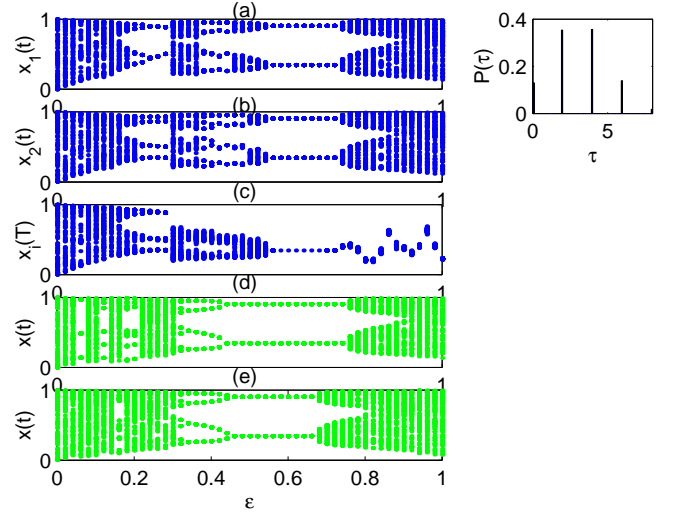}}
\caption{As Fig. 7 but with all-even delays.}
\end{figure}

Above a certain coupling strength the array synchronizes in a single
cluster that displays either steady-state or time-dependent dynamics
[in Figs. 7(c)-9(c) there is a single cloud of points for
$\epsilon>\approx 0.5$]; below this coupling strength the array
splits into clusters and the elements of each cluster evolve along
similar time-dependent orbits (the bifurcation diagrams for the
elements $i=1$ and $i=2$ of the array are similar, even for low
$\epsilon$).

For a map with self-feedback loops with ``mixed'' delays, the fixed
point is stabilized for increasing $\epsilon$ after a
period-doubling bifurcation [Fig. 7(d),(e)]. In contrast, for
feedback loops with all-odd delays the fixed point is stabilized
after a Hopf bifurcation [Fig. 8(d),(e)]. For all-even delays the
fixed point is not stable for any $\epsilon$ [but the period-two
orbit is stable in a certain range of $\epsilon$, Fig. 9(d),(e)].
These results are in agreement with the analysis of the previous
section, where we found that for all-odd delays the fixed point
changes stability when a pair of complex eigenvalues cross the unit
circle, and for all-even delays the fixed point changes stability
when a real eigenvalue crosses the unit circle at $\lambda=-1$. The
bifurcation diagrams of the $i=1$ and $i=2$ maps of the array
display similar features [Figs. 7-9(a), 7-9(b)]: the fixed point is
stable in a range of $\epsilon$ for ``mixed'' and all-odd delays,
while the period-two orbit is stable in a range of $\epsilon$ in the
case of all-even delays.

Exponentially distributed delays [$\tau_{ij} = \tau_0 + \mathrm{Int}
(c \xi)$, where $\xi$ is exponentially distributed, positive, with
unit mean] yield similar bifurcation diagrams, shown in Fig. 10.
Furthermore, the instability scenario for fixed $\epsilon$ and
increasing $a$ (i.e., a scan along a vertical line in Fig. 4) is
also very similar, as shown in Fig. 11.
\begin{figure}
\center
\resizebox{1.0\columnwidth}{!}{\includegraphics{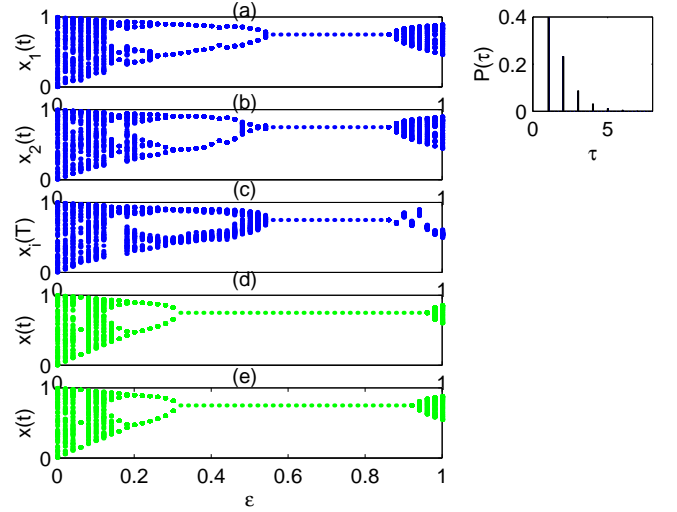}}
\caption{As Fig. 7 but with exponentially distributed delays
($\tau_0=1$, $c=1$).}
\end{figure}
\begin{figure}
\center
\resizebox{1.0\columnwidth}{!}{\includegraphics{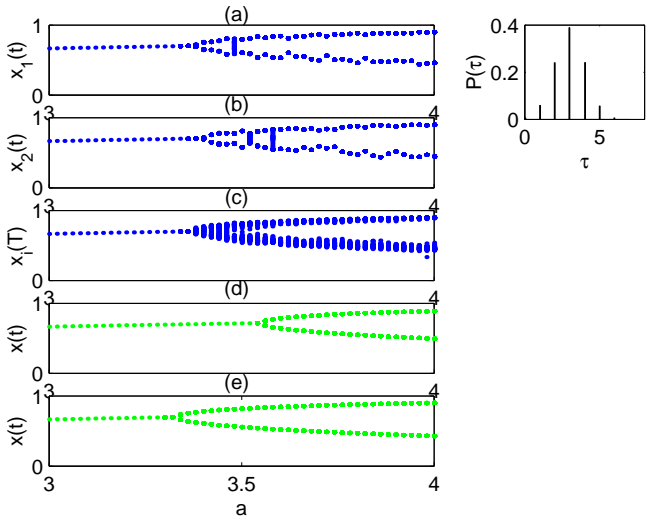}}
\caption{As Fig. 7 but varying $a$ while keeping $\epsilon=0.3$
fixed.}
\end{figure}

\section{Discussion}

The similarities between a map of the array and a map with
self-feedback loops can be interpreted in the framework of the
analogy between globally coupled maps (with instantaneous coupling),
and a single map subjected to an external drive, studied by Cosenza
and Parravano in Refs.\cite{cosenza1,cosenza2}. The authors
considered
\begin{equation}
\label{cosenza}
 x_i(t+1)= (1-\epsilon) f[x_i(t)] + \epsilon H(x_1(t) \dots x_N(t)),
\end{equation}
where $H$ is a global coupling function that is invariant to
argument permutations [$H(x_1, \dots, x_i, \dots, x_j, \dots
x_N)=H(x_1, \dots, x_j, \dots, x_i, \dots, x_N)$ $\forall$ $i$ and
$j$], and showed that the clustering behavior of the array can be
analyzed through the analogy with the driven map,
\begin{equation}
\label{cosenza2}
 x(t+1)= (1-\epsilon) f[x(t)] + \epsilon F(t),
\end{equation}
where $F(t)$ is a external forcing (assumed to be periodic). The
analogy holds because in Eq.(\ref{cosenza}) all the elements of the
array are affected by the coupling function $H$ in exactly the same
way at all times, and therefore the behavior of any element of the
array is equivalent to the behavior of the driven map, Eq.
(\ref{cosenza2}).

To analyze whether this analogy can be extended to the case of {\it
delayed}-coupling, we calculated the mean field coupling term at
site $i$ of the array,
\begin{equation}
\label{mean_field}
 H_i(t)=  \frac {1}{N}  \sum_{j=1}^N f[x_j(t-\tau_{ij})],
\end{equation}
and compared with the driving term of the map,
\begin{equation}
F(t) = \frac {1}{N} \sum_{j=1}^N f[x(t-\tau_{j})].
\end{equation}
We found that for $N$ large, $H_i(t)$ is nearly the same for all the
elements of the array, regardless of the array dynamics, and its
time evolution is similar to that of the driving term of the map,
$F$. As an example, Fig. 12 displays the mean field coupling term
and the driving term, for parameters corresponding to the
bifurcation diagrams shown in Fig. 7. In Fig. 12(a) we plot 100
consecutive values of the mean field at one element of the array,
$H_1$ (after transients die away). Figure 12(b) displays the mean
field at all elements, $H_i$ with $i=1\dots N$, at time $t=T$ (large
enough to let transients die away). It can be seen that $H_i \simeq
H_j$ even for low values of $\epsilon$. Figure 12(c) displays 100
consecutive values of the driving term of the map with feedback
loops, $F(t)$ (after transients die away), and it is observed that
above a certain value of $\epsilon$ ($\epsilon \approx 0.2$) $F
\simeq H_1$.
\begin{figure}
\center
\resizebox{0.70\columnwidth}{!}{\includegraphics{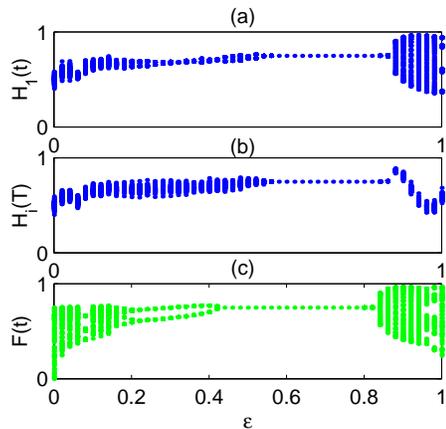}}
\caption{(a) Mean field for the $i=1$ map of the array vs.
$\epsilon$. (b) Mean field for each map of the array, $H_i$ with
$i=1\dots N$, vs. $\epsilon$. (c) Driving term of the map with
feedback loops, $F$, vs. $\epsilon$. $N=100$, $a=4$, the delays are
the same as in Fig. 7.}
\end{figure}

We speculate that the analogy with the single map has its roots in
the fact that the elements of the array display similar temporal
variation, i.e., they evolve along equal (or similar) orbits, even
when the array splits into clusters. Therefore, a map of the array
''perceives'' signals coming from other maps as nearly
indistinguishable from signals coming from self-feedback loops. This
can also be thought as an ergodic property of the dynamics, since
the average over the ensemble, $H_i$, is nearly equal to the average
over time, $F$.

The analogy holds even if the delays are all equal, $c=0$, as shown
in Fig. 13. Above a certain coupling strength ($\epsilon \approx
0.45$) the array synchronizes in-phase, $x_i(t)=x_j(t)$ $\forall$
$i$, $j$, and therefore the analogy is mathematically trivial since
in the synchronization manifold the evolution equation of one
element of the array and the evolution equation for a map with a
single self-feedback loop are exactly the same. However,
multistability in the delayed map (i.e., the coexistence of
different stable orbits) might lead to competition phenomena in the
array of coupled maps, as different elements might tend to evolve
along different orbits, depending on the initial conditions. The
investigation of this type of dynamics is the object of future work.
\begin{figure}
\center
\resizebox{0.70\columnwidth}{!}{\includegraphics{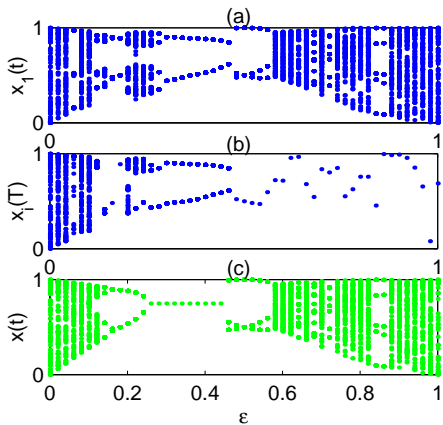}}
\caption{As Fig. 7 but with fixed delays ($\tau_0=1$, $c=0$).}
\end{figure}

\section{Summary and Conclusions}

We investigated the dynamics of mutually coupled logistic maps
focusing on the influence of the delay times of the interactions
between the maps, and comparing with the dynamics of a map with
several time-delayed self-feedback loops. By using some mathematical
tools such as the Gershgorin theorem we derived analytic stability
and instability conditions for the fixed point solution of the map
with feedback loops. We found that the stabilization of the array in
the fixed-point solution can be well understood in terms of the
dynamics of the map. Specifically, for randomly distributed delay
times, if $N$ and $\epsilon$ are large enough the fixed-point is
stable for both, the map with self-feedback loops and the array.
Also, if the delay times are all even, for both, the single map with
$N$ delayed loops and the array of $N$ delayed-coupled maps, we
observed that the stability region of the fixed-point is reduced to
the ``trivial'' region ($a \le 3$) regardless of the coupling
strength. The results presented here provide another example of an
ensemble of mutually coupled interacting units, where understanding
the dynamics of a single unit with self-feedback loops is relevant
for understanding the macroscopic behavior of the ensemble.
\section{Acknowledgments}
CM acknowledges support from the ``Ramon and Cajal'' Program (MCyT
of Spain).
\section{Appendix}
In this appendix we demonstrate Eq. (\ref{polinomio}). We start from
Eq. (\ref{a0}),
\begin{eqnarray}
\label{a0_apendix}
 [(1-\epsilon)f'(x_{0})+ \alpha_0 -\lambda] \det C - \det B=0,
\end{eqnarray}
and calculate the determinant of matrix $C$ recursively:
\begin{eqnarray}
\det C = \det\begin{pmatrix}
  -\lambda & 0 & 0& \hdots &0  & 0 \\
 1  &-\lambda & 0 & \hdots &0 & 0 \\
 0 & 1 & -\lambda & \hdots &0 & 0 \\
\vdots & & & &\vdots \\
 0& 0& 0 & \hdots &-\lambda & 0 \\
 0& 0& 0& \hdots & 1 & -\lambda
\end{pmatrix}\nonumber \\
= -\lambda \det\begin{pmatrix}
 -\lambda & 0 & \hdots &0 & 0 \\
 1 & -\lambda & \hdots &0 & 0 \\
\vdots & & & &\vdots \\
 0& 0 & \hdots &-\lambda & 0 \\
 0& 0& \hdots & 1 & -\lambda
\end{pmatrix} \nonumber \\
-\det\begin{pmatrix}
0 & 0& \hdots &0  & 0 \\
1 & -\lambda & \hdots &0 & 0 \\
\vdots & & & &\vdots \\
0& 0 & \hdots &-\lambda & 0 \\
0& 0& \hdots & 1 & -\lambda
\end{pmatrix}
\end{eqnarray}
The second determinant is zero because the elements of the first row
are all zero. We obtain
\begin{equation}
\label{detC} \det C(M) = -\lambda \det C(M-1) = \dots =
(-\lambda)^M.
\end{equation}
The determinant of matrix $B$ can also be calculated recursively:
\begin{eqnarray}
\det B =\det \begin{pmatrix}  \alpha_{1}  & \alpha_{2} & \alpha_{3}
& \hdots & \alpha_{M-1}&
 \alpha_{M} \\
 1  &-\lambda & 0& \hdots &0 & 0 \\
 0 & 1 & -\lambda & \hdots &0 & 0 \\
\vdots & & & &\vdots \\
 0& 0& 0&\hdots &-\lambda & 0 \\
 0& 0& 0&\hdots & 1 & -\lambda
\end{pmatrix} \nonumber \\
=\alpha_{1}\det
\begin{pmatrix}
 -\lambda & 0& \hdots &0 & 0 \\
 1 & -\lambda & \hdots &0 & 0 \\
\vdots & & & &\vdots \\
 0& 0&\hdots &-\lambda & 0 \\
 0& 0&\hdots & 1 & -\lambda
\end{pmatrix} \nonumber \\
- \det
\begin{pmatrix}
\alpha_{2} & \alpha_{3} & \hdots & \alpha_{M-1}&
 \alpha_{M} \\
 1 & -\lambda & \hdots &0 & 0 \\
\vdots & & & &\vdots \\
0& 0&\hdots &-\lambda & 0 \\
0& 0&\hdots & 1 & -\lambda
\end{pmatrix}
\end{eqnarray}
It can be noticed that the first matrix is $C(M-1)$ while the
determinant of the second matrix can be calculated recursively,
\begin{eqnarray}
\label{detB} \det B(M) = \alpha_{1} \det C(M-1) -\det B(M-1),
\end{eqnarray}
where
\begin{eqnarray}
B(M-1) =
\begin{pmatrix}  \alpha_{2} & \hdots & \alpha_{M-1}&
 \alpha_{M} \\
\vdots & & &\vdots \\
0&\hdots &-\lambda & 0 \\
0&\hdots & 1 & -\lambda
\end{pmatrix}.
\end{eqnarray}
Substituting in (\ref{a0_apendix})
\begin{eqnarray}
[(1-\epsilon)f'(x_{0})+ \alpha_0 &-&\lambda] \det C(M) - [\alpha_{1} \det
C(M-1) \nonumber\\ - \alpha_{2} \det C(M-2) &+& \det B(M-2)] =0
\end{eqnarray}
Using Eqs.(\ref{detC}) and(\ref{detB}) we obtain
\begin{eqnarray}
[(1-\epsilon)f'(x_{0})+ \alpha_0&-&\lambda] (-\lambda)^M - \alpha_{1}
(-\lambda)^{M-1} \nonumber\\ + \alpha_{2} (-\lambda)^{M-2} &-& \dots
=0
\end{eqnarray}
which gives
\begin{eqnarray}
(-\lambda)^{M+1} +[(1-\epsilon) f'(x_{0})+\alpha_0](-\lambda)^M
\nonumber\\+\sum_{j=1}^M (-1)^j \alpha_{j} (-\lambda)^{M-j} =0,
\end{eqnarray}
that can be simplified to
\begin{eqnarray}
-\lambda^{M+1} +[(1-\epsilon) f'(x_{0})+ \alpha_0]\lambda^M +\sum_{j=1}^M
\alpha_{j} \lambda^{M-j} =0.
\end{eqnarray}

\end{document}